\documentclass[a4paper,preprintnumbers,showpacs,onecolumn,superscriptaddress,nofootinbib,amsmath,amssymb,notitlepage,bibtem]{revtex4-1}

\usepackage{enumitem}
\usepackage{times}
\usepackage{dcolumn}
\usepackage{mathrsfs}
\usepackage{comment}
\usepackage{hyperref}
\usepackage{amsmath}
\usepackage{mathtools}
\usepackage{bbm}
\usepackage{bm}
\usepackage{amsfonts}
\usepackage{amssymb}
\usepackage{mathrsfs}
\usepackage[scr=rsfs,cal=boondox]{mathalfa}
\usepackage{wasysym}
\usepackage{graphicx}
\usepackage[]{subfigure}
\usepackage{filecontents}
\usepackage[dvipsnames]{xcolor}
\usepackage{bbold}
\usepackage[scr=rsfs,cal=boondox]{mathalfa}
\usepackage{stackengine}
\stackMath
\newcommand\tsup[2][2]{%
 \def\useanchorwidth{T}%
  \ifnum#1>1%
    \stackon[-1.3ex]{\tsup[\numexpr#1-1\relax]{#2}}{\mathchar"307E}%
  \else%
    \stackon[-1ex]{#2}{\mathchar"307E}%
  \fi%
}

\hypersetup{
    bookmarks=true,         
    unicode=true,          
    pdftoolbar=true,        
   pdfmenubar=true,        
    pdffitwindow=true,     
   pdfstartview={FitH},     
   pdftitle={My title},     
    pdfauthor={author},      
    pdfsubject={Subject},    
    pdfcreator={Creator},    
    pdfproducer={Producer},  
   pdfkeywords={keyword1} 
    pdfnewwindow=true,       
    colorlinks=true ,        
    linkcolor=Blue  ,        
    citecolor=Blue,      
    filecolor=Blue,       
    urlcolor=Blue            
}

\begin{document}

\title{
Cosmic slowing down of acceleration with the Chaplygin-Jacobi gas as a dark fluid?}
\author{J.A.S.Fortunato}
\email{jeferson.fortunato@edu.ufes.br}
\affiliation{PPGCosmo, CCE, Universidade Federal do Esp\'irito Santo, Av. Fernando Ferrari, 540, CEP 29.075-910, Vit\'oria, ES, Brazil}
\affiliation{N\'ucleo Cosmo-UFES, CCE, Universidade Federal do Esp\'{\i}rito Santo, Av. Fernando Ferrari, 540, CEP 29.075-910, Vit\'oria, ES, Brazil}
\affiliation{Instituto Argentino de Radioastronom\'{\i}a, C.C. No. 5, 1894 Buenos Aires, Argentina}

\author{W.S. Hip\'olito-Ricaldi}
\email{wiliam.ricaldi@ufes.br}
\affiliation{Departamento de Ci\^encias Naturais, CEUNES, Universidade Federal do Esp\'{\i}rito Santo, Rodovia BR 101 Norte, km. 60,CEP 29.932-540, S\~ao Mateus, ES, Brazil}
\affiliation{N\'ucleo Cosmo-UFES, CCE, Universidade Federal do Esp\'{\i}rito Santo, Av. Fernando Ferrari, 540, CEP 29.075-910, Vit\'oria, ES, Brazil}

\author{N. Videla}
\email{nelson.videla@pucv.cl}
\affiliation{Instituto de F\'isica,  Pontificia Universidad Cat\'olica de Valpara\'iso, Avenida Universidad 331, Curauma, Valpara\'iso, Chile}

\author{J.R. Villanueva}
\email{jose.villanueva@uv.cl}
\affiliation{Instituto de F\'{i}sica y Astronom\'{i}a, Facultad de Ciencias, Universidad de Valpara\'{i}so,\\
Avenida Gran Breta\~{n}a 1111, Playa Ancha, Valpara\'{i}so, Chile}


\begin{abstract}

{A particular generalization of the Chaplygin inflationary model, using the formalism of Hamilton-Jacobi and elliptic functions, results in a more general non-linear Chaplygin-type equation of state (Chaplygin-Jacobi model). We investigate the implementation of this model as a dark energy (DE) fluid to explain the recent acceleration of the universe. Unlike $\Lambda$CDM and other Chaplygin-like fluids, where the final fate of the universe is an eternal de Sitter (dS) phase, the dynamics of this model allows for the possibility of a decelerating phase in the future, following the current accelerating phase. In other words, a transient acceleration  arises, accounting for the recently claimed slowing down phenomenon. This Chaplygin-Jacobi model shows important differences compared to the standard and generalized Chaplygin gas models. Additionally, we perform a Markov Chain Monte Carlo (MCMC) analysis using several datasets, including Type Ia Supernovae (SNIa), Cosmic Chronometers (CC), Fast Radio Bursts (FRBs), and Baryon Acoustic Oscillations (BAO) to examine the observational viability of the model. Our results indicate that, although a transient phase of accelerated expansion is supported by current observations in the context of the Chaplygin-Jacobi model, this model is strongly disfavored in comparison with $\Lambda$CDM.}

\bigskip

\end{abstract}

\maketitle

\section{Introduction}
One of the greatest challenges in modern science is understanding the universe, particularly its origin, evolution, and final fate. In this regard, the hot big bang (HBB) cosmology, based on four-dimensional General Relativity (GR) \cite{Einstein:1916vd}, has been widely accepted as the standard paradigm describing how the universe expanded from an initial singularity. During the longest part of its lifetime, the universe has undergone a decelerating expansion, being dominated first by radiation and then by matter. However, the cosmic history  includes two phases of accelerated expansion: one at very early times and another at late times. The first accelerating phase corresponds to inflation, which supplies an explanation of the observed large scale structure (LSS),  as it provides the primordial, almost adiabatic and scale-invariant perturbations that undergo gravitational collapse to form galaxies and clusters of galaxies \cite{Abazajian:2013vfg}. The second accelerating phase corresponds to the current cosmic acceleration, which is supported from a large number of observational  evidence \cite{Riess1998,Perlmutter1998,WMAP2003,Aghanim2020,cooke2018one,Eisenstein2005}, which also indicates that our universe is spatially almost flat, and currently dominated by dark energy (DE) and cold dark matter (DM). The accelerated expansion of the present universe is attributed to DE, which is an exotic component having negative pressure, such as the cosmological constant  $\Lambda$ \cite{Padmanabhan:2002ji,Sahni:1999gb,Carroll:2000fy}. The $\Lambda$CDM model, has proven to be successful in explaining a wide range of cosmological observations \cite{Riess1998,Perlmutter1998,WMAP2003,Aghanim2020,cooke2018one,Eisenstein2005}. Despite its success, it faces several challenges, namely: (i) the cosmological constant problem \cite{Weinberg1989,Martin:2012bt},  (ii) the cosmic coincidence problem (or the so-called “why now”
problem) \cite{Zlatev1998,Arkani-Hamed:2000ifx}, and more recently, (iii) the tension between measurements of the Hubble parameter $H_0$  \cite{DiValentino2021,Riess:2020fzl} and (iv) the $S_8$ tension ($S_8=\sigma_8 \sqrt{ \Omega_m/0.3}$)   \cite{DiValentino2020}. 

Several approaches have been proposed to address $\Lambda$CDM problems. They generally fall into two groups: one modifies GR, while the other, dynamical DE, involves models where the properties of DE vary over time unlike in the $\Lambda$CDM.  Several candidates for dynamical DE have been studied in recent years including the Chaplygin gas (CG). For a review of DE models, see Refs. \cite{Copeland:2006wr,Nojiri:2006ri,Clifton:2011jh, Bamba2012,Joyce:2016vqv}. The CG's equation of state (EoS) has a connection with string and brane theories \cite{Bilic:2001cg,Heydari-Fard:2007vcg} and it can admit a supersymmetric generalization \cite{Jackiw2000}. Chaplygin cosmology can be understood as the study of the dark sector through a CG fluid that satisfies the relation $p=-\frac{B}{\rho}$ \cite{Kamenshchik:2001cp,Fabris:2002xx,Gorini:2002kf}, and its generalizations, like the Generalized Chaplygin Gas (GCG) with an equation of state $p=-\frac{B}{\rho^{\alpha}}$ \cite{Bilic:2001cg,Bento:2002ps}, where $p$ represents pressure (assumed to be negative to produce an accelerated stage at late times), $\rho$ is the energy density, and $B$ is a positive constant (to ensure $p<0)$ and $0<\alpha\leq 1$. These models belong to the unified models class, which provide a description of the matter-dominated era and the late accelerating era with only one component rather than two (for other unified models, see e.g. \cite{Hipolito2009,Hipolito2010,Zimdahl2011}). However, observationally, the CG model has shown problems with instabilities in their corresponding trajectories \cite{Perrotta2004} and when tested with various observations, such as Type Ia Supernovae (SNIa) , X-ray gas mass fraction of clusters, and Hubble rate-redshift data \cite{biesiada2005,Colistete:2005yx,Zhu:2004aq,Wu:2006pe,Makler:2002jv}. Moreover, although the GCG model is in agreement with background observational data \cite{biesiada2005,Colistete:2005yx,Zhu:2004aq,Wu:2006pe,Makler:2002jv,delCampo:2009cz}, it suffers from an unexpected blow-up in the matter power spectrum caused by adiabatic pressure perturbations \cite{Sandvik:2002jz}. 
This undesired effect can be avoided if some kind of non-adiabaticity is introduced in the model (see e.g. \cite{vomMarttens:2017cuz}). Moreover, in recent years, several GCG modifications or generalizations have been proposed to study the dark sector. Among others, some of them are the modified Chaplygin gas (MCG) \cite{Benaoum2022,Debnath2004}, the new generalized Chaplygin gas (NCG) \cite{Zhang2004},  and viscous generalized Chaplygin gas (VGCG) \cite{Zhai2005,Hernandez-Almada:2021osl}. For recent observational results involving those models, see e.g. \cite{Zheng2022}. 

Recently, a further generalization has been proposed through the Hamilton-Jacobi formalism in the inflationary context by means of elliptic functions \cite{Rengo1}. This leads to the development of a more general Chaplygin-like EoS, which we will refer to as Chaplygin-Jacobi Gas model (CJG). The CJG EoS is expressed as follows:

\begin{equation}
\label{prro}
p=-\frac{B\,k}{\rho^{\alpha}}-k'\rho\left(2-\frac{1}{B}\rho^{\alpha+1}\right)\,.
\end{equation}
In the above equation, $k$ represents the modulus of the elliptic function, and $k'=1-k$ is the complementary modulus. It is noteworthy that the GCG is obtained with $k=1$ and $B>0$.  In recent years, the CJG has been studied in contexts different from this work \cite{Rengo1,Rengo2,Rengo3,Debnath:2021pxy,Mukherjee2023,Chaudhary2023l,Rengo4}.

Considering that Chaplygin-like fluids offer an interesting framework to study phenomenology beyond the $\Lambda$CDM, the main goal of the present work is to explore the viability of the CJG for describing the late universe. First, we integrate the balance equation and find its solutions for the energy density, pressure and   EoS parameter. Then, we perform an analysis of its parameter space to find the region in which physical solutions are possible. After obtaining an analytical solution for the Hubble rate, and in order to compare the background evolution of the CJG with that of the $\Lambda$CDM model, we introduce parameters where derivatives of the scale factor beyond the second-order appear. To this end, one option is to study the so-called statefinder parameters, $r$ and $s$, defined as follows \cite{Sahni:2002fz,Alam:2003sc}.
\begin{eqnarray}
r & \equiv & \frac{\dddot{a}}{a H^3}, \\
s & \equiv & \frac{r-1}{3 (q-\frac{1}{2})} \label{rsz},
\end{eqnarray}
where the dot denotes differentiation with respect to cosmic time $t$, $H=\dot{a}/a$ represents the Hubble rate, and $q=-\ddot{a}/(a H^2)$ denotes the deceleration parameter. It is worth noting that the statefinder parameters involve third derivatives of the scale factor with respect to cosmic time, in contrast to the Hubble rate and the deceleration parameter, which are expressed in terms of the first and second time derivatives of the scale factor, respectively. Statefinder analysis serves as a diagnostic tool for understanding the dynamics of the universe's expansion. These parameters are computed for our specific DE model, and as we will discuss later, $r$ and $s$ can differ significantly from $\Lambda$CDM even if they predict very similar expansion histories.

While the emergent cosmology may not realize a unified model like most of the CG-like models, it offers a phenomenology with interesting results distinct from those of the CG model and its generalizations, particularly in future times. As is expected for any DE model, a fluid satisfying the relation (\ref{prro}) is subdominant in a multi-component universe during early times. This accurately reproduces a radiation-dominated era followed by a matter-dominated era, resulting in a decelerating universe. Subsequently, the DE fluid dominates, leading to an accelerating phase. However, a specific region in the parameter space permits a transient acceleration-deceleration in the future, indicating the possibility of a decelerating phase following the current accelerating phase. This phenomenon is commonly referred to as the \textit{cosmic slowing down} of the current acceleration. DE models with a constant EoS like $\Lambda$CDM or the $w$CDM cannot exhibit the slowing down feature \cite{Zhang2018}, as they predict a de Sitter (dS) phase as the final fate of the universe. However, depending on the set of parameters, some DE models, especially those with dynamical EoS or parametrizations like the Chevallier-Polarski-Linder (CPL) model and others, allow for this possibility (see e.g. \cite{Vargas2011,Hu2015,Magana2014,Zhang2018,Bolotin2020}). Here, we demonstrate that a universe with a fluid satisfying the relation (\ref{prro}) as DE can also exhibit the slowing down behavior. Additionally, we conduct a data analysis to investigate whether recent astronomical and cosmological observations support this transient behavior in the context of the CJG. To achieve this, we utilize the latest supernova  Pantheon + data (PAN$^+$), Cosmic Chronometers data (CC) and BAO data. Furthermore, we incorporate observations from Fast Radio Bursts (FRBs) into our tests, as they have recently been shown to provide an interesting complement to other cosmological probes.

Our work is organized as follows: after this introduction, in Section \ref{two}, we derive solutions for the energy density, pressure, and EoS parameter. Next, we conduct an analysis of the parameter space to determine the sub-regions where physical solutions are feasible and the sub-regions where the slowing down phenomenon appears. In Section \ref{three}, we delve into the background cosmological dynamics of the CJG in a multifluid context. This includes obtaining an analytical solution for the Hubble rate $H(z)$ as a function of redshift and introducing the deceleration and Statefinder parameters tailored for the CJG. In Section \ref{four}, we present the data and relevant equations for the implementation of the  MCMC analysis. In Section \ref{results}, we present the results of the MCMC analysis and  evaluate numerically the cosmic evolution of the CJG against redshift, comparing these results with those of the $\Lambda$CDM model. Finally, we summarize our findings and present our conclusions in Section \ref{conclu}. Throughout our work, we adopt the mostly positive metric signature $(-,+,+,+)$ and utilize natural units where $c=\hbar=1$.

\section{Chaplygin-Jacobi dark energy} \label{two}
As mentioned before, the Hamilton-Jacobi formalism allows for obtaining a particular generalization of the Chaplygin inflationary model through elliptical functions, resulting in a more general Chaplygin-like EoS parameter \cite{Rengo1}
\begin{equation}
\label{omega} \omega= -\frac{B\,k}{\rho^{\alpha+1}}-2k'+\frac{k'}{B}\rho^{\alpha+1},
\end{equation}
where $k'=1-k$, and $\alpha$, $B$ and $k$ are constants with $0<k<1$ and $0<k'<1$. In principle, the EoS in (\ref{prro}) reduces to a GCG case when $k=1$, $B>0$ and $0<\alpha<1$. The case where $B>0$, $k=1$ and $\alpha=0$ reproduces $\Lambda$CDM. There is no other case in which one can obtain the MGC, NCG, or another Chaplygin-like EoS. In the GCG context, only the first term in Eq. (\ref{prro}) appears.  Then, for  $B > 0$ and $0 <\alpha < 1$, the pressure is always negative, allowing an explanation for  the late accelerated expansion of the universe.  However, the CJG has two additional contributions to the pressure. The second term of Eq. (\ref{prro}) is always negative, but the first and  third terms always have opposite signs. For instance,  for $B>0$, the first term is negative and the third term is positive, while for $B<0$  the behaviour is reversed. Eventually, depending on the parameter space, the positive term might dominate and the pressure might  become less negative opening the  the possibility of a slowing down of the cosmic acceleration. This becomes more evident when  looking  at the energy density and pressure as functions of the scale factor. Let us first consider one universe dominated by a fluid with an EoS as in Eq. (\ref{prro}) and solve the energy balance equation.

In the context of a universe described 
by the Friedman-Lema\^{\i}tre-Robertson-Walker (FLRW) metric, the energy balance equation  for an EoS given by  Eq. (\ref{prro}) reads
\begin{eqnarray}
\frac{\dot{\rho}}{\rho}=-3 \frac{\dot{a}}{a}\left(-\frac{B\,k}{\rho^{\alpha+1}}+2k-1+\frac{(1-k)}{B}\rho^{\alpha+1}\right)=-3 \frac{\dot{a}}{a}\left(\frac{-B k-(1-2k)\rho^{\alpha+1}+
\frac{1-k}{B}\rho^{2(\alpha+1)}}
{\rho^{\alpha+1}}\right),
\end{eqnarray}
where $a$ is the scale factor of the homogeneous and isotropic metric, and a dot denotes differentiation with respect to cosmic time. This equation can be integrated directly resulting in
\begin{eqnarray} \label{rhodim}
 \rho(a)=\left[B-\frac{B A}{A k'-a^{3(1+\alpha)}}\right]^{\frac{1}{1+\alpha}}  \,,
\end{eqnarray}
where $A$ is an integration constant. 
{It should be noted that a particular choice of parameters $B$, $k$ and $\alpha$ does not
uniquely determine the CJG model, as it is also necessary to specify  the value of $\rho$ at a particular redshift.
This can be addressed by introducing $B = \rho_0^{\left(
1+\alpha\right)}B_s$, where $\rho_0 = \rho (a_0)>0$ is the current density. This ensures that the choice of
$B_s$, $k$ and $\alpha$  uniquely specifies the CJG model, and Eq.~(\ref{rhodim}) can be conveniently rewritten as}
\begin{eqnarray}
 \rho(a)=\rho_0\left[B_s-\frac{B_s A}{A k'-a^{3(1+\alpha)}}  \right]^{\frac{1}{1+\alpha}}\,,
\end{eqnarray}
{where} the integration constant $A$ is determined by the condition  $\rho=\rho_0$ at the present time, so that
\begin{eqnarray}
 \label{contA} A= \frac{1-B_s}{1-k(1-B_s)},
\end{eqnarray}
and thus,
\begin{eqnarray}\label{rho}
 \rho(a)=\rho_o \left[B_s-\frac{B_s (1-B_s)}{(1-B_s) k'-(k'+B_s\,k)a^{3(1+\alpha)}}  \right]^{\frac{1}{1+\alpha}}\,.
\end{eqnarray}
Notice that the energy density spans a three-dimensional parameter space. Accordingly, the pressure of the CJG as a function of the scale factor can be computed by replacing Eq.~(\ref{rho}) into Eq.~(\ref{prro}). Although, in principle, solution (\ref{rho}) has a  three-dimensional parameter space
\begin{equation}
{\cal{M}}=\{(\alpha, B_s, k)/ \,\alpha \in \mathbb{R}, B_s\in \mathbb{R}, 0<k<1\},
\end{equation}
some regions of this space can lead to singularities, divergences in the past, or negative densities, resulting in  non-physical solutions across the entire space  ${\cal{M}}$. However, certain sub-regions of  ${\cal{M}}$ allow solutions without these problems. Let us  find the sub-regions in ${\cal{M}}$ where we can have physical solutions for the model. 
\subsection{Physical solutions}
From a physical standpoint,  we  impose that the energy density must always be positive and that there are no singularities within the interval $0 < a < 1$.  For convenience, it is better to analyze two different regimes for the parameter  $\alpha$: (i) $\alpha + 1 <0$ and (ii) $\alpha +1\ge 0$. Let us analyze each case separately
\subsubsection{Case  $1+\alpha<0$}
{For  $1+\alpha=-|1+\alpha| <0$, the energy density in Eq.~(\ref{rho}) can be rewritten as 
\begin{eqnarray}\label{rhoneg}
\left(\frac{\rho}{\rho_0}\right)^{|1+\alpha|}=\frac{1}{B_s}
\frac{1+(B_s-1)(k+k' a^{3|1+\alpha|})}
{
1+(B_s-1) k (1- a^{3|1+\alpha|})} \,.
\end{eqnarray}}
In that case, a singularity happens for

\begin{eqnarray}\label{condition1}
a^{3|1+\alpha|}_c = 1 + \frac{1}{k(B_s-1)}.
\end{eqnarray}
To avoid any singularity in  $0<a<1$, $a_c$ must be outside of this interval. One possibility is to have $a^{3|\alpha+1|}_c \in \mathbb{R}$ and $a_c > 1 $. This condition  is possible only if the second term in Eq.~(\ref{condition1}) is always positive, i.e.,  for $B_s>1$. Therefore, we have the first sub-region with physical solutions
\begin{eqnarray}
{\cal{N}_1}=\{(\alpha, B_s, k)/ \alpha < -1, B_s>1 , 0<k<1\} \,.
\end{eqnarray}
Another  possibility is that {$a^{3|\alpha+1|}_c \notin \mathbb{R}$}  and $a_c<0$. This occurs only when  $-\frac{k'}{k}<B_s<1$. Nevertheless, in the past, when $a \ll 1$, we have the limit

\begin{eqnarray} \label{limit1}
\left(\frac{\rho}{\rho_0}\right)^{|1+\alpha|}\rightarrow\frac{1}{B_s}\,,
\end{eqnarray}
so, we discard negatives values for $B_s$,  identifying another sub-region ${\cal{N}_2} \subset {\cal{M}}$ with physical solutions,  such that
\begin{eqnarray}
{\cal{N}_2}=\{(\alpha, B_s, k)/ \alpha < -1, 0<B_s<1 , 0<k<1\} \,.
\end{eqnarray}
\subsubsection{Case $\alpha+1\ge0$}
In the case where $\alpha +1\ge 0$, Eq. (\ref{rho}) has a singularity at
\begin{eqnarray}\label{singa}
a_c^{3(1+\alpha)}= 1-\frac{B_s}{k'+B_sk}\,.
\end{eqnarray}
As in the previous case, to avoid a singularity in  $0<a<1$, we have the conditions (i) $a_c >1$ and (ii) $a_c<0$. In each case,  we  divide the analysis into two parts: for $B_s>0$ and for $B_s<0$. In condition (i), the second term on the right side of  Eq.~(\ref{singa}) must be negative, which for $B_s>0$, only occurs when $k'+B_sk <0$. However, this condition never holds when  $B_s>0$.  On the other hand, for $B_s <0$, the second term on the right side of Eq.~(\ref{singa}) is negative when $k'+B_sk >0$, which is true if $-\frac{k'}{k}<B_S<0$. Consequently, we identify another sub-region ${\cal{N}_3} \subset {\cal{M}}$ with physical solutions, such that 
\begin{eqnarray}
{\cal{N}}_3=\{(\alpha, B_s, k)/ \alpha \ge  -1, -\frac{k'}{k}< B_s < 0, 0<k<1\}.
\end{eqnarray}
At this point, it is worth mentioning that $B_s$ being negative is not a problem. Unlike in other Chaplygin-like cosmologies where the first term in Eq.~(\ref{prro}) has to be negative, in the CJG context this term can be positive while the third term is negative. The three terms together can ensure the negativity of the entire expression in Eq.~(\ref{prro}). Additionally, notice that for early times, i.e., $a \ll 1$, we have that
\begin{eqnarray} \label{limit2}
\rho\rightarrow \rho_{i}\equiv \rho_0
 \left( \frac{-B_s\,k}{k'}\right)^{\frac{1}{1+\alpha}}, 
\end{eqnarray}
which is perfectly compatible with negative values for $B_s$. Moreover, for $B_s > 0$, condition (ii) is satisfied when $0 < k' + B_s k < B_s$, which is the case for  $B_s > 1$. However, positive values are not allowed by the limit in Eq.~(\ref{limit2}). Finally, for $B_s <0 $, condition (ii) is satisfied when $B_s < k' + B_s k <0$, which is valid for $B_s<-\frac{k'}{k}$. Therefore, we also have physical solutions in the sub-region ${\cal{N}}_4 \subset  {\cal{M}}$ such that
\begin{eqnarray}
{\cal{N}}_4=\{(\alpha, B_s, k)/ \alpha > -1,  B_s < -\frac{k'}{k}, 0<k<1\}.
\end{eqnarray}
Summarizing, we have physical solutions in 
\begin{eqnarray}
{\cal{N}}=  {\cal{N}}_1 \cup {\cal{N}}_2 \cup {\cal{N}}_3 \cup {\cal{N}}_4 \,.
\end{eqnarray}
We remark that the density is always positive in ${\cal{N}}$. Additionally,  unlike  CG, GCG, MCG, NCG, and VCG, the limit in the past for CJG (Eqs.~(\ref{limit1}) and (\ref{limit2})) does not correspond to a matter component. This implies that  CJG cannot be used as a unified model.
\subsection{Conditions for cosmic slowing down}
The cosmic acceleration is usually attributed to the dominance of a negative pressure component. For instance, in the $\Lambda$CDM case, where $p=-\rho$, the pressure is always negative and acceleration will extend forever, driving the universe towards a dS phase as a final state. The same fate is predicted by GCG and other Chaplygin-like models. However, for the CJG model, depending on contributions to the pressure, this can be less negative, thereby opening the possibility of a decelerating phase immediately after the present accelerating phase. Consequently, it is interesting to check the existence of transient acceleration in the sub-region ${\cal{N}}$. To this end, we need to start with the following expressions for the deceleration parameter and for the relation between the Hubble function and a general EoS parameter $w(z)$
\begin{eqnarray} \label{dece}
q(z) = -1 +\frac{1+z}{E}\frac{d E}{d z} \,, \qquad E^2(z) \propto \text{exp} \left[3\int^z_0 \frac{1+w(z')}{1+z'}dz'\right] \,.
\end{eqnarray}
where $E(z)=H(z)/H_0$ and $H = \frac{\dot{a}}{a}$ is the Hubble rate. A transition between accelerated and decelerated phases requires a switch to a negative slope in the deceleration parameter, i.e., $\frac{dq}{dz} < 0$. Using Eqs.~(\ref{dece}), it is possible to express this condition in terms of the EoS parameter (see \cite{Zhang2018} for details). This leads to the conclusion that slowing down acceleration requires at least the following condition to be met \cite{Zhang2018}:  
\begin{eqnarray} \label{condition}
\frac{d w}{d z} < 0 \,.
\end{eqnarray}
Clearly, the $\Lambda$CDM and $w$CDM models do not predict cosmic slowing down \cite{Zhang2018}. Similarly, the GCG, MCG, and NGC models also predict a final dS phase for the universe, unlike the CJG model. Thus, we need to determine under which conditions a different fate from the dS phase will occur in the studied model. 

The  EoS parameter $w$  as  a function of the scale parameter  can be found using Eqs.~(\ref{omega}) and (\ref{rho}), which results in
\begin{eqnarray}
w = -1 + \frac{(1-B_s)k}{k(1-B_s)+(k'+B_s k)a^{3(1+\alpha)}} + \frac{(1-B_s)k'}{(k'+B_s k)a^{3(1+\alpha)}-(1-B_s)k'} \,.
\end{eqnarray}
In the appendix, we present a detailed evaluation of the condition in Eq.~(\ref{condition}). The results indicate that cosmic slowing down is possible in ${\cal{N}}_1$ and ${\cal{N}}_4$. 


 \section{Cosmology of the CJG} \label{three}
Here we study a cosmological model based on the results from the previous section using the CJG fluid as the DE component. First, we assume the cosmic substratum is dynamically dominated by a mixture of radiation ($r$), pressureless dark matter ($m$) and a DE component ($de$). Radiation and matter are modeled by barotropic fluids with EoS parameters $w_r=1/3$ and $w_m \ll 1$, respectively. The DE energy component is modeled as a CJG fluid  with EoS (\ref{prro}).

In General Relativity, the Friedmann equation for a homogeneous, isotropic, and spatially flat three-component universe is 
\begin{eqnarray}
3H^2=8\pi\,G (\rho_r+\rho_m+\rho_{de}) \,.
\end{eqnarray}
Here, $\rho_r$, $\rho_m$, and $\rho_{de}$ are the radiation, pressureless dark matter, and DE densities, respectively. Moreover, energy conservation is satisfied by all the components separately
\begin{eqnarray} \label{conservation}
\dot{\rho}_i + 3H(\rho_i+p_i) =0 \,,
\end{eqnarray}
where $i=r,m,de$. As it is well-known, solutions of Eq. (\ref{conservation}) for radiation and matter are found to be $\rho_r =\rho_{r0}a^{-4}$ and $\rho_m =\rho_{m0}a^{-3}$, respectively. For DE, the solution is given by Eq. (\ref{rho}). The sub-index “0” denotes the current value of any given quantity. 

\begin{figure*} [!h]
\includegraphics[width=0.48\linewidth]{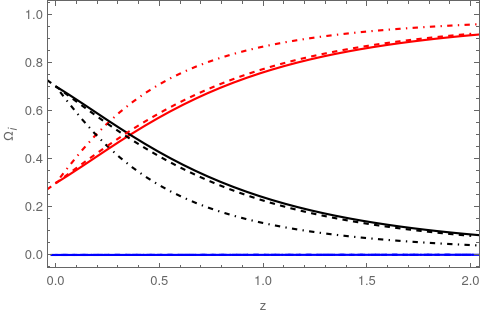} \qquad 
\includegraphics[width=0.48\linewidth]{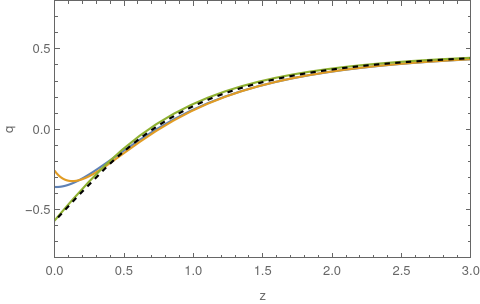} 
\caption{Fractional densities for radiation, matter and DE (left) and  deceleration parameter (right) as functions of  redshift for different sets of parameters in the sub-region ${\cal{N}}$ (described in the text) for CJG and $\Lambda$CDM cosmologies.}
\label{Fig1}
\end{figure*}

The past evolution is restricted by the necessity of a radiation and  matter-dominated epochs to guarantee, for instance, the Big Bang Nucleosynthesis (BBN) and  cosmic structure formation. To be a viable DE model, the CJG must be subdominant in the past, reproducing correctly a radiation domination era and subsequently, the matter domination era giving a decelerating universe. In recent times,  the DE fluid must dominate to give rise  the accelerating stage of the universe. A fluid with EoS given by  (\ref{prro}) and within a subspace of parameters $\cal{N}$ reproduces all these points properly, as  can be observed in Fig.~\ref{Fig1} (left side), where the fractional densities are shown as a function of redshift. In all cases, the blue curves represent radiation, the red ones represent matter, and the black curves represent DE fluid.  For the dot-dashed curve, parameter values are  $\Omega_{m0}=0.30$, $\alpha=0.1$, $B_s=-0.6$, $k=0.43$, corresponding to a point in ${\cal{N}}_3$. For the solid line, we used $\Omega_{m0}=0.30$, $\alpha=1.3$, $B_s=-1.4$, $k=0.46$, corresponding to a point in ${\cal{N}}_4$. 
For the sake of comparison, we also plotted the corresponding functions in the  $\Lambda$CDM model  using the same colored dashed curves.

On the other hand, the Hubble rate can be written as a function of the scale factor $a$. However, it will  be more useful to have the dimensionless Hubble rate as a function of $z$
\begin{eqnarray} \label{hubblez}
E^2(z)=\Omega_{r0}(1+z)^4+\Omega_{m0}(1+z)^3+(1-\Omega_{m0}-\Omega_{r0})\left[B_s+\frac{B_s (1-B_s)(1+z)^{3(1+\alpha)}}{(k'+B_s\,k)-(1-B_s)k'(1+z)^{3(1+\alpha)}}  \right]^{\frac{1}{1+\alpha}},
\end{eqnarray}
where we have used the fact that each energy density is usually expressed in terms of the dimensionless density parameter, defined as $\Omega_i=\rho_i/\rho_{cr}$ with $\rho_{cr}=3H^2/(8\pi G)$ being the critical density. 

Having obtained the analytical expression for $E(z)$, we may proceed to analyze our model using the statefinder diagnostic. It will be useful to express thee Statefinder parameters in terms of the dimensionless Hubble rate and the redshift $z$. The cosmic times derivatives are written as redshift derivatives according to
\begin{equation}\nonumber
\frac{d}{dt}=\frac{d}{dz}\frac{dz}{da}\frac{da}{dt}=-(1+z)H(z)\frac{d}{dz}\,.
\end{equation}
In this way, the set of Statefinder parameters becomes
\begin{align}\label{rz} 
    r&=1+\frac{1}{2E^2}\left[(1+z)^2\frac{d^2E^2}{dz^2}-2(1+z)\frac{dE^2}{dz}\right],\\ 
    s&=\frac{1}{3} \frac{(1+z)^2\frac{d^2E^2}{dz^2}-2(1+z)\frac{dE^2}{dz}}{(1+z)\frac{dE^2}{dz}-3E^2}\label{sz}.
\end{align}
Accordingly, upon replacement of Eq. (\ref{hubblez}) into Eqs. (\ref{rz})-(\ref{sz}), the specific expressions for the Statefinder parameters are obtained (not shown), and they will be plotted in Section \ref{results} using the best-fit values. Furthermore, we compute the effective EoS parameter $w_{eff}$ for our CJG cosmological model, which encodes information about the universe's composition, the evolution of its energy density and the dynamics of its expansion. The general expression for $w_{eff}$ depends on the first derivative of $E^2(z)$ with respect to the redshift $z$ as follows
\begin{eqnarray}
    w_{eff}=-1+\frac{1+z}{3E^2}\frac{d E^2}{dz}.\label{weff}
\end{eqnarray}
Thus, by replacing Eq.(\ref{hubblez}) into (\ref{weff}), $w_{eff}$ for our specific CJG model is found (not shown).

As discussed above, for a universe dominated only by a CJG, a particular region in the subspace of parameters allows for transient acceleration-deceleration in the future. This feature must remain when a more realistic model of the universe is considered, including radiation and dark matter. Therefore, the future cosmological evolution within the framework of CJG may be very different from a dS. This is shown in Fig.~\ref{Fig1} (right side), where several curves for deceleration parameter as a function of redshift were plotted for different sets of parameters. The dashed dark curve on the right side of Fig.~\ref{Fig1} represents the case for  $\Lambda$CDM, in which, as it is known, the accelerated expansion will continue forever.  The green curve, which corresponds to the set $\Omega_m = 0.3$, $\alpha = 1.1$, $B_s=-0.81$ and $k=0.55$, also indicates an eternal accelerated expansion. However, the last two curves, the blue one for $\Omega_m = 0.3$, $\alpha = 0.1$, $B_s=-1.5$, and $k=0.43$, and the orange one for $\Omega_m = 0.3$, $\alpha = 0.8$, $B_s=-1.41$, and $k=0.42$,  exhibit a transition between an accelerated  and  a decelerated phase in the future. In all the curves, the transition from a decelerated evolution during the matter-dominated epoch to an accelerating phase in the past can be observed. 

The possibility of future transient acceleration is not new. In the past, evidence was found for a slowing down of the expansion rate of the universe, or equivalently, for an increase in the deceleration parameter at redshifts close to the present epoch $z \approx 0$ \cite{Shafieloo2009,Guimaraes2011,Cai2011,Costa2010, Fabris2009,Vargas2011,Zimdahl2013}. More recently, this possibility has been explored in light of recent data \cite{Magana2014,Shahalam2015,Hu2015,Bolotin2020,Escobal2023}. In fact, over the last few years, several authors have discussed the possibility that the accelerated expansion might be a transient phenomenon, i.e., that there might be a transition back to decelerated expansion \cite{Albrecht1999,Barrow2000,Bento2001}.

So far, we have explored the concept of cosmic slowing down with a CJG fluid from a theoretical standpoint. However, in recent years, a large amount of cosmological data has emerged enabling very precise statistical tests. In the next section, we will use different probes to study the viability of a CJG as dark energy and what the data can tell us about the cosmic slowing down in the CJG framework. In this regard, several quantities describing the background evolution, such as the Hubble rate, the deceleration parameter, and the statefinder parameters, will be evaluated after obtaining the best-fit values from the aforementioned observational analysis. 
\section{Data} \label{four}
In Section \ref{two}, the regions in the parameter space where the model has physical solutions were limited by imposing physical conditions on the fluid density (\ref{rho}). In this section, we will utilize current observational data to estimate the free parameters and their associated error bars. Through this process, we aim to determine the region in the parameter space where a model based on CJG could be compatible with observations. We will utilize cosmic chronometers (CC), Type Ia Supernovae (SNIa),  Fast Gamma Ray Bursts (FRBs), and Baryon Acoustic Oscillations (BAO) datasets, along with their joint analysis, to perform a MCMC study. Implementations were carried out using the publicly available Python package \verb|Polychord| \citep[][]{handley2015polychord}. Now, let us briefly describe each of the datasets used.
\subsection{Cosmic Chronometers}
The Hubble rate is the quantity that most directly characterizes the expansion of the universe. Over the last few years, its measurement has advanced significantly, and currently, there are two well-known methods to obtain $H(z)$ data. The first method utilizes the differential age of galaxies and is referred to as cosmic chronometers (CC) \cite{Jimenez2002,Stern2010,Zhang2014,Moresco2016}. The second method measures the peaks of the baryon acoustic oscillations (BAO) observing the typical acoustic scale in the line-of-sight direction \cite{Gaztanaga2008,BOSS:2012gof,BOSS2013,BOSS2014}. The CC technique relies on measuring the age difference between two ensembles of passively evolving massive galaxies at slightly different redshifts, from which one can determine the derivative of redshift with respect to cosmic time \cite{Jimenez2002}. This method has the advantage of avoiding systematic errors that arise when measuring the absolute ages of individual galaxies, instead allowing the measurement of the relative age difference between them. In a homogeneous and isotropic universe, the Hubble rate and the derivative of redshift with respect to cosmic time are related by
\begin{equation}
    H(z)=-\frac{1}{1+z}\frac{dz}{dt}\simeq-\frac{1}{1+z}\frac{\Delta z}{\Delta t}.
\end{equation}

Since the CC method does not rely on a particular functional form of the expansion history or spatial geometry, it can be regarded as a model-independent method. On the other hand, the BAO approach is based on the relation between $H(z)$ and the comoving differential radial distance. This method requires knowledge of the comoving BAO scale ($r_{\text{BAO}}$), which is derived from CMB measurements. This fact makes this probe not entirely cosmology-independent, since typically, in the derivation of the sound horizon scale from CMB a cosmological model is assumed \cite{Jiao2022}. Therefore, in this work, we have chosen to use the 31 measurements of the Hubble parameter $H(z)$ via the CC method, compiled in \cite{wang2020reconstructing}.

To estimate the region of parameters compatible with the CC dataset, we need to evaluate the likelihood function

\begin{eqnarray}
    -2ln(\mathcal{L_{\rm CC}})=\chi^2=\Delta \Vec{C}^T C^{-1}_{\rm CC}\Delta \Vec{C},
\end{eqnarray}

where the vector $\Delta \Vec{C}$ is the difference between $H_i$, i.e. each of the individual measurements of the sample considered, and $H^{\rm theo}$, the set of Hubble parameters calculated from Eq. (\ref{hubblez}). The inverse covariance matrix is given by $C^{-1}_{CC}$. Here, we employ the method introduced by \cite{kazantzidis2018evolution} to simulate possible correlations among these data points. This involves introducing positive correlations to 20\% of the dataset by randomly selecting off-diagonal elements into the covariance matrix while maintaining its symmetry. The magnitude of these elements is set to

\begin{equation}
C_{ij}=0.5\sigma_i\sigma_j,    
\end{equation}

being $0.5\sigma_i\sigma_j$ the $1\sigma$ errors of the elements $i$ and $j$.
\subsection{SNIa}
Type Ia supernovae (SNIa) constitute a distinct and significant class of stellar phenomena, emerging during the terminal phases of stellar evolution 
and culminating in a catastrophic explosion that disperses stellar material into the cosmos. Given that SNIa are exceptionally bright, they can be detected at large cosmological distances and are sufficiently common to be found in large numbers. In fact, they have been crucial for mapping the cosmic expansion of the universe. The evidence for the accelerated cosmic expansion came from SNIa data analysis \cite{Riess1998,Perlmutter1998}.

For the cosmological analysis, the SNIa light curves must be standardized \cite{amendola2010dark} to correct variations in brightness and other factors. This standardization ensures that their luminosities can be used as reliable distance indicators through the distance modulus equation \cite{brout2022pantheon+}

\begin{equation}
    \mu =m_b+\alpha x_1 - \beta c - M_B, 
\end{equation}
where $\alpha$ and $\beta$ are parameters that relate stretch $x_1$ and color $c$ to luminosity. $M_B$ and $m_b$ are the fiducial magnitude and the light-curve amplitude of the supernova, respectively. With this data, the $\chi^2$ method is applied:

\begin{eqnarray}
    -2ln(\mathcal{L_{\rm SNIa}})=\chi^2=\Delta \Vec{D}^T C^{-1}_{stat+syst}\Delta \Vec{D},
\end{eqnarray}
being $C^{-1}_{stat+syst}$ the combined statistical and systematic covariance matrix, and $\Vec{D}$, the vector related to the distance module residuals computed from

\begin{eqnarray}
    \Delta D_ i = \mu_ i - \mu_{model}(z_i)\,.
\end{eqnarray}

Here, $\mu_{model}(z_i) = 5log(d_L(z_i)/10 ~\rm pc)$ is the theoretical distance modulus estimated from the host galaxy redshift and the luminosity distance $d_L$, which considering a spatially flat universe is given by \begin{equation}
    d_L(z) = (1+z)~c\int_0^z\frac{z^\prime}{H(z^\prime)}.
\end{equation}
In this paper, we constrain the CJG cosmological parameters using the Pantheon+ dataset (PAN$^+$), which comprises 1624 data points (when decoupled from the 77 SNe in Cepheid hosts used in SH0ES collaboration to calibrate the absolute magnitude). 

\subsection{FRBs}
Fast Radio Bursts (FRBs) are intense pulses in the radio spectrum originating from cosmological distances, characterized by an extremely short duration (a few milliseconds). In recent years, they have emerged as interesting complementary cosmological probes. Discovered in 2007 \cite{Lorimer2007}, hundreds of bursts have been reported so far. While their progenitors remain unknown (for some progenitor models, see e.g., \cite{Bochenek2020,Zhang2020,Bhandari2020}), 24 of them have been localized, allowing determination of their host galaxies and redshifts. In this study, we perform our calculations on a subset of these FRBs, compiled by \cite{yang2022finding}, excluding the nearest FRB, named FRB200110E, as it carries little cosmological information. Given the current uncertainties in modeling the extragalactic components of pulses, certain parameter choices can result in non-physical negative values for the dispersion measure DM. Consequently, we exclude FRB181030 (\(z = 0.0039\)) and FRB210807 (\(z = 0.12927\)) from our sample, as their DMs are estimated to be negative based on the methodology outlined below \cite{Fortunato2024}.

An FRB pulse is dispersed by the intergalactic medium during its path to the observer, generating a time delay between different radio frequencies that compose the observed signal. This dispersion is quantified by the dispersion measure $\mathrm{DM}$, which is related to the column density of free electrons $n_e$ along the FRB line of sight $l$, weighted by redshift as 

\begin{equation}
    \mathrm{DM} = \int \frac{n_e}{(1+z)}dl.
\end{equation}

From an observational point of view, the observed dispersion measure $\rm DM_{obs}$, is expected to have four different contributions, such that
\begin{equation}\label{dmobs}
    \rm DM_{obs} = \mathrm{DM}_{\mathrm{ISM}}+\mathrm{DM}_{\mathrm{halo}} + \mathrm{DM}_{\mathrm{IGM}}+\frac{\mathrm{DM_{\mathrm{host}}}}{(1+z)}\,.
\end{equation}
The first two terms come from the intergalactic medium, and the last two come from the extragalactic medium. $\mathrm{DM}_{\mathrm{ISM}}$  corresponds to the Milky Way interstellar medium contribution,  $\mathrm{DM}_{\mathrm{halo}}$ is related to the Milky Way galactic halo,   $\mathrm{DM}_{\mathrm{IGM}}$ is the contribution from the intergalactic medium ($\mathrm{IGM}$), which incorporates the cosmological dependence, and $\mathrm{DM}_{\mathrm{host}}$ is the host galaxy component corrected with $(1+z)^{-1}$ to account for cosmological expansion for a given FRB source at redshift $z$. $\mathrm{DM}_{\mathrm{ISM}}$ is calculated using galactic electron distribution models ($\mathrm{NE2001}$ \cite{cordes2002ne2001} or $\mathrm{YMW16}$ \cite{yao2017new}). Since recent works have claimed that the YMW16 model may overestimate $\mathrm{DM}_{\mathrm{ISM}}$ at low Galactic latitudes \cite{koch2021}, we use the NE2001 approach. $\mathrm{DM}_{\mathrm{halo}}$ has been estimated to be in the interval $50 < \mathrm{DM}_{\mathrm{halo}} < 100~\mathrm{pc}.\mathrm{cm}^{-3}$ \cite{prochaska2019probing}. However, to be conservative, we assume $\mathrm{DM}_{\mathrm{halo}} = 50~\mathrm{pc}.\mathrm{cm}^{-3}$ as in, for example, \cite{macquart2020census}.

The exact characteristics of the host environment and the exact contribution of $\rm DM_{\rm host}$ for each FRB are still uncertain. Therefore, the estimation of $\rm DM_{\rm host}$ considers a probabilistic approach. Following \cite{hagstotz2022new}, we use the stochastic distribution
 \begin{equation}
    P(\mathrm{DM}_\mathrm{host}) = N \big( \langle \mathrm{DM}_\mathrm{host} \rangle, \sigma_\mathrm{host}^2 \big)  ,
\end{equation}
where $N$ is a normal distribution with a mean value $\langle \mathrm{DM}_\mathrm{host} \rangle = 100 (1+z_\mathrm{host})^{-1}  \mathrm{pc}.\mathrm{cm}^{-3}$ and variance $\sigma_\mathrm{host} = 50 (1+z_\mathrm{host})^{-1}  \mathrm{pc}. \mathrm{cm}^{-3}$. On the other hand, the dominant contribution in Eq.(\ref{dmobs}) is due to $\mathrm{DM}_{\mathrm{IGM}}$. Recently, cosmological simulations have shown that the distribution of ${\rm DM_{IGM}}$ is influenced by the distribution of baryons around galactic halos and the number of collapsed structures intersecting a given line of sight. Two different approaches have been used to take this influence into account: the non-Gaussian \cite{prochaska2019frbs,yang2022finding,wu2020new} and the Gaussian \cite{jaroszynski2019fast, hagstotz2022new, zhang2023cosmology} approaches. In this work, we use the Gaussian approach; however, recent studies with the 23 localized FRB data points have shown that there is no appreciable difference between using either approaches \cite{Fortunato2023}. The Gaussian approach assumes a normal distribution around the mean
\begin{equation}\label{DM_igm}
{\rm \langle DM_{IGM} \rangle } = \frac{3c\, {\rm \Omega_b} H_0 }{8\pi Gm_{\rm p}}\int^z_0\frac{(1+z^\prime)f_{\rm IGM}(z^\prime)f_{\rm e}(z^\prime)}{E(z^\prime)}dz^\prime,
\end{equation}
with standard deviation interpolated in the range $\sigma_{\rm IGM}(z=0)\approx 10 ~\mathrm{pc}.\mathrm{cm}^{-3}$ and $\sigma_{\rm IGM}(z=1)\approx 400 ~\mathrm{pc}.\mathrm{cm}^{-3}$. In Eq. (\ref{DM_igm}), $f_{\rm e} (z) = Y_{\rm H}X_{\rm e,H}(z)+\frac{1}{2}Y_{\rm He}X_{\rm e,He}(z)$. The cosmic baryon density, the proton mass, and the fraction of baryon mass in the IGM are denoted as $\Omega_b$, $m_{\rm p}$, and $f_{\rm IGM}$, respectively. Because both hydrogen and helium are completely ionized at $z<3$, the ionization fractions of each species are $X_{\rm e,H}=X_{\rm e,He}=1$. Additionally, we consider an IGM with a hydrogen mass fraction $Y_{\rm H}=0.75$ and a helium mass fraction $Y_{\rm He}=0.25$. Moreover, several analyses have found a constant value for the fraction of baryon mass, $f_{\rm IGM} = 0.82$ (see e.g., \cite{shull2012baryon,Fortunato2023}). Finally, in Eq. (\ref{dmobs}), $\mathrm{DM}_{\mathrm{IGM}}$ is estimated by $\mathrm{DM}_{\rm IGM} = \rm DM_{\rm obs} - \rm DM_{\rm local} - \rm DM_{\rm host}(1+z)^{-1}$, with uncertainty given as
\begin{equation}\label{error}
    \sigma_{\rm IGM}(z)=\sqrt{\sigma_{\rm obs}(z)^2+\sigma^2_{\rm local}+\left(\frac{\sigma_{\rm host}(z)}{1+z}\right)^2},
\end{equation}
where $\sigma_{\rm obs}$ and $\sigma_{\rm host}$ are the errors related to $\rm DM_{\rm obs}$ and $\rm DM_{\rm host}$, respectively. Meanwhile, $\sigma_{\rm local}$ is the sum of uncertainties in $\rm DM_{\rm ISM}$ and $\rm DM_{\rm halo}$. Following \cite{hagstotz2022new}, we approximate $\sigma_{\rm local} \approx 30~\rm{pc.cm}^{-3}$.

In order to compute the likelihood, we must consider that every observed dispersion measure $\rm DM_i$ at a given redshift $z_i$ is related to a Gaussian individual likelihood through
       \begin{equation}
            \mathcal{L}(\rm{DM}_i, z_i) = \frac{1}{\sqrt{2 \pi \sigma_i^2}} \exp \left[ \frac{\bigl(\rm{DM}_i - \rm{DM}^{\rm{theo}}(z_i) \bigr)^2} {2 \sigma_i^2} \right] \,,
        \end{equation}
        where $\rm{DM}^{\rm theo}(z_i)$ is the theoretical contribution computed as $        \rm{DM}^{\rm{theo}}(z_i) =   \rm DM_{obs} - \rm DM_{\rm ISM}-\rm{DM}_{halo} 
        =  \rm{DM}_{\rm{IGM}} (z_i) +    \rm{DM}_{\rm{host}} (z_i) $, with  
        \begin{equation}
            \sigma^2(z_i) =  \sigma_{\rm{MW}}^2 +       \sigma_{\rm{host}}^2(z_i) +       \sigma_{\rm{IGM}}^2(z_i) \, .
        \end{equation}
        Considering all events as independent, the total likelihood of the sample is 
        \begin{equation}
        \label{eq:likelihood_tot}
        \mathcal{L}_{\rm{FRBs}} = \prod_i  \mathcal{L}_i \: ,
        \end{equation}
        and the computation of the product is executed for every the 23 FRB data points.
\subsection{BAO}
A key cosmological probe for investigating the LSS of the universe corresponds to the Baryonic Acoustic Oscillations (BAO). These oscillations, arising from waves in the photon-baryon fluid of the early universe, created distinct density fluctuations. As the universe expanded and cooled, these fluctuations left an imprint as a characteristic scale of approximately $150~\rm Mpc$ in the matter distribution, most notably in the separation between galaxies. This scale serves as a "standard ruler" for measuring cosmic distances and the expansion rate of the universe.

BAO data is derived by analyzing the large-scale distribution of galaxy clusters, with measurements taken both transversely and radially to determine the angular diameter distance and the Hubble rate across various redshifts. These measurements play a significant role in constraining the expansion history of the universe. Large-scale surveys, such as the Sloan Digital Sky Survey (SDSS) \cite{accetta2022seventeenth}, have mapped BAO signatures with remarkable precision. 

To compare cosmological models with observational data, we consider the following key distance ratios:

\begin{equation}
    \frac{d_m(z)}{r_d} = \frac{d_l(z)}{r_d(1+z)},
\end{equation}

\noindent being \( d_m(z) \) is the transverse comoving distance, and \( r_d \) represents the sound horizon at the drag epoch. Throughout this analysis, we adopt the value of \( r_d \) determined by the latest Planck collaboration results \cite{Aghanim2020}. Another fundamental quantity is the ratio 

\begin{equation}
    \frac{d_H(z)}{r_d} = \frac{c}{r_d H(z)},
\end{equation}

\noindent where \( d_H(z) \) is the Hubble distance and \( H(z) \) denotes the Hubble parameter. Finally, we consider the volume-averaged angular diameter distance, given by

\begin{equation}
    \frac{d_v(z)}{r_d} = \frac{[z d_m^2(z) d_H(z)]^{1/3}}{r_d}.
\end{equation}

\noindent This set of distance measures provides essential constraints for testing cosmological models against large-scale structure and BAO observations.

In this study, we use the most up-to-date release from the DESI collaboration, in which they provide BAO measurements combining DESI 2024 \cite{adame2024desi} galaxy and quasar BAO measurements with SDSS-BOSS and eBOSS results, to estimate the parameters of the models investigated in this work. In this case, the likelihood for each ratio $X = d_m/r_D, ~d_H/r_D, ~d_v/r_D$ is given by

\begin{eqnarray}
    -2ln(\mathcal{L_{\rm BAO}})=\chi_{\rm BAO}^2=\sum^N_i\Big\{\Big[\frac{X_i-X_{theo}}{\sigma_{X_i}} \Big]^2+\ln(2\pi\sigma^2_{X_i})\Big\}.
\end{eqnarray}


\section{Analyses and Results}\label{results}
We have implemented two separate scenarios for the MCMC joint analysis: (a) CC + FRBs + BAO, which entirely excludes SNIa, and (b) CC + FRBs + BAO + $\rm PAN^{+}$. In order to perform all analyses, we  fixed $\Omega_{r}= 5.38 \times 10^{-5}$ to be consistent with CMB observations \cite{ParticleDataGroup:2020ssz}, and considered the following set of free parameters: $H_0$, $\Omega_ m$, $\alpha$, $B_s$, and $k$. For consistency, we used the same flat priors in all cases, such as $[20, 100]$ for $H_0$ and $[0, 1]$ for $\Omega_m$. For $\alpha$, $B_s$, and $k$, we suitably considered the sub-regions in ${\cal{N}}$. Moreover, when the additional parameter $M_B$ is introduced, we used a flat prior of $[-20, -18]$.

\begin{figure}[h!]
    \includegraphics[width=15cm]{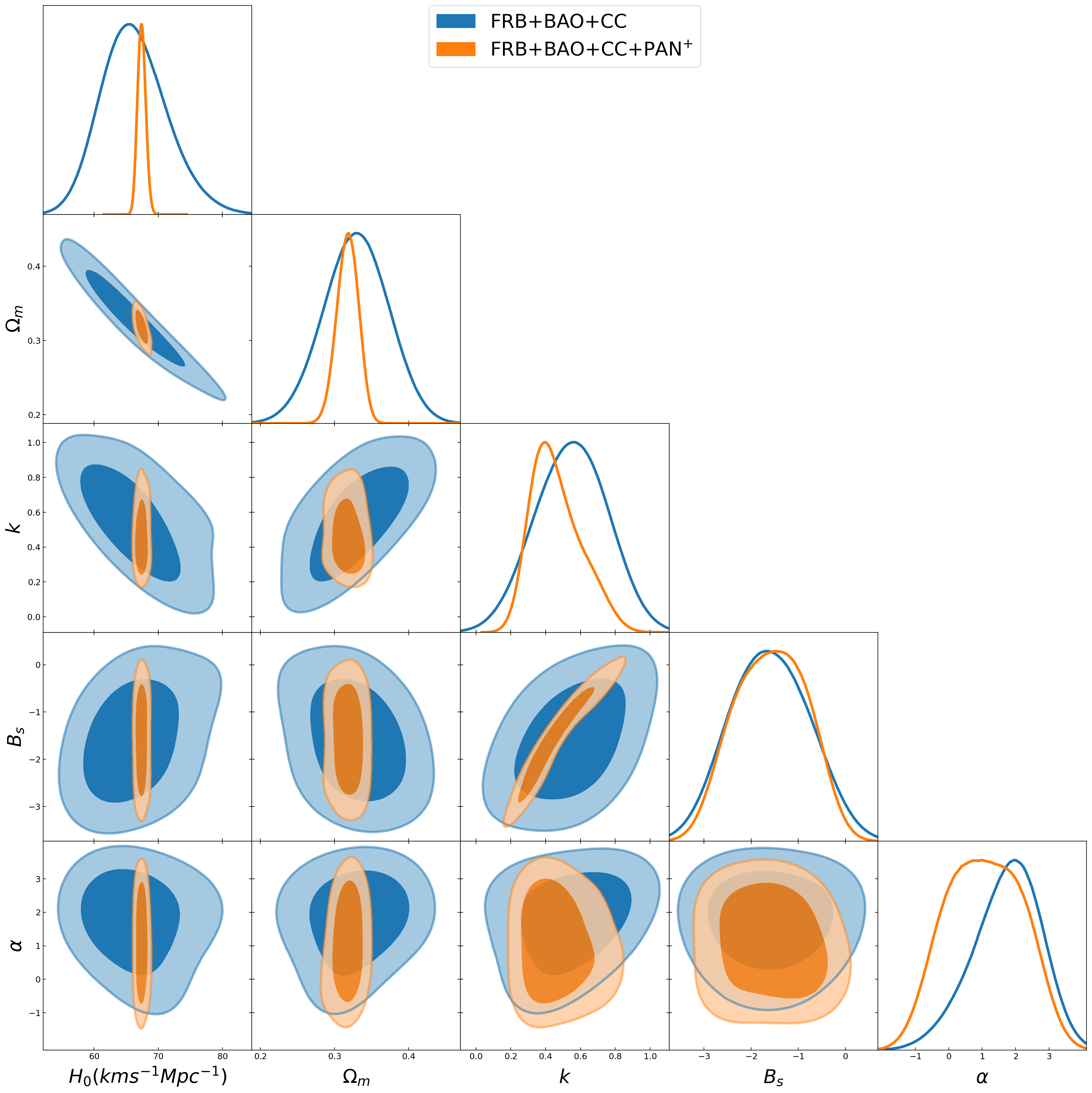}
    \caption{1$\sigma$ and 2$\sigma$ C.L. curves from the MCMC analysis for CJG using (a) FRBs + BAO + CC and (b) FRBs + BAO + CC + PAN$^+$.}
    \label{Fig2}
\end{figure}

Results for the (a) and (b) analyses are presented in Fig.~\ref{Fig2} and Table~\ref{table1}. Both analyses exhibit good agreement within the 1$\sigma$ statistical confidence level (C.L.), indicating concordance between the CJG cosmology in the region ${\cal{N}}_3 \cup {\cal{N}}_4$ and observations. Their best-fit values support a slowing-down phase. In the contour plots of Fig.~\ref{Fig2}, it is possible to observe some general features. For example, in general, regions with $1+\alpha > 0$ and $B_s < 0$ are preferred by observations; i.e. the data favor the $\cal{N}_3 \cup \cal{N}_4$ region, with a slight trend towards $\cal{N}_3$. This can be seen in Fig.~\ref{Ez} (left), where the plane $B_s$-$k$  is shown together with the curve $B_s=-\frac{k'}{k}$ (black dashed curve) separating $\cal{N}_3$ (de Sitter future) and $\cal{N}_4$ (non-de Sitter future). Notice that the best-fit point indicated by $\mathrm{x}$ lies within the slowdown region. Moreover, there is a positive correlation between the $B_s$ and $k$ parameters, which can be explained by the  $k$-dependence of the intervals for $B_s$  (see Sec. \ref{two}). Additionally, in Table \ref{table1}, we present the $\chi^2_{min}$ and $\chi^2_\nu$ values, along with the Akaike Information Criterion (AIC) and Bayesian Information Criterion (BIC) values. The $\chi^2_{\nu} \sim 1$ for CJG in cases (a) and (b) indicates that the data are well fitted by the model. 

With the best-fit values obtained from the joint analysis CC+FRBs+BAO+$\rm PAN^{+}$ for example, we can plot relevant quantities that describe the cosmic evolution at the background level and compare them with those from the $\Lambda$CDM model. {Fig.~\ref{Ez} (right) shows the plot of the relative deviation between the Hubble function of CJG and that of the $\Lambda$CDM model, $1-H_{CJG}(z)/H_{\Lambda \text{CDM}}$. As shown from Fig.\ref{Ez}, $H(z)$ becomes indistinguishable for both models during the past, while it approaches very close values in the present time, resulting in a very small negative relative difference. In future times, $H(z)$ for CJG becomes smaller than that for the $\Lambda$CDM model, leading to a positive relative deviation. Conversely, cosmic acceleration is accounted for by the deceleration parameter, which is depicted in Fig.~\ref{figure4}. As can be seen, the best-fit values (black curves) support cosmic deceleration in the future. These findings are interesting because they present a different behavior  compared to the $\Lambda$CDM case (blue line on the left side of Fig.\ref{figure4}) and other types of Chaplygin-like fluids such as NCG, MCG, and GCG, as shown on the right side of Fig.\ref{figure4}\footnote{To plot the deceleration curve for Chaplygin-like models, we used the best-fit values for NCG, MCG, and GCG found in \cite{Zheng2022}}. Nevertheless, it is important to note that the 1$\sigma$ region (gray region in Fig.~\ref{figure4}) is relatively large and sufficiently inclusive to also encompass a dS final fate for the universe. In general, further data and analyses are needed to refine our understanding of the final fate of the universe

\begin{table*}[h!]
\centering
\resizebox{\columnwidth}{!}{
\begin{tabular}{cccccccccccc}
\hline
\hline
Model & Data & $H_0$ & $\Omega_m$ & $B_s$ & $k$ & $\alpha$ & $M_B$ & $\chi^2_{\text{min}}$ & $\chi^2_{\nu}$ & AIC & BIC \\
\hline
CJG & (a) & $66.10^{+4.20}_{-5.40}$ & $0.329\pm0.041$ & $-1.60\pm0.79$ & $0.55\pm0.20$ & $1.73^{+1.10}_{-0.75}$ & - & 55.11 & 1.04 & 67.11 & 79.57 \\
     & (b) & $67.41\pm0.62$ & $0.318\pm0.014$ & $-1.57\pm0.75$ & $0.46^{+0.08}_{-0.18}$ & $1.1\pm1.1$ & $-19.423^{+0.018}_{-0.020}$ & 1623.85 & 0.97 & 1637.85 & 1675.83 \\
\hline
$\Lambda$CDM & (a) & $68.66\pm0.89$ & $0.306\pm0.016$ & - & - & - & - & 48.67 & 0.87 & 60.67 & 60.90 \\
             & (b) & $67.56\pm0.57$ & $0.3296\pm0.0097$ & - & - & - & $-19.429\pm0.016$ & 1595.39 & 0.95 & 1607.39 & 1625.09 \\
\hline
\end{tabular}}
\vspace{-.1cm}
\caption{Best-fit values and their 1$\sigma$ errors for (a) FRB + BAO + CC and (b) FRB + BAO + CC + $\rm PAN^{+}$, for the CJG and $\Lambda$CDM models. We also include the $\chi^2_{\text{min}}$, $\chi^2_{\nu}$, AIC, and BIC values.}
\label{table1}
\end{table*}



\begin{figure*}[h!]
\includegraphics[width=0.49\linewidth]{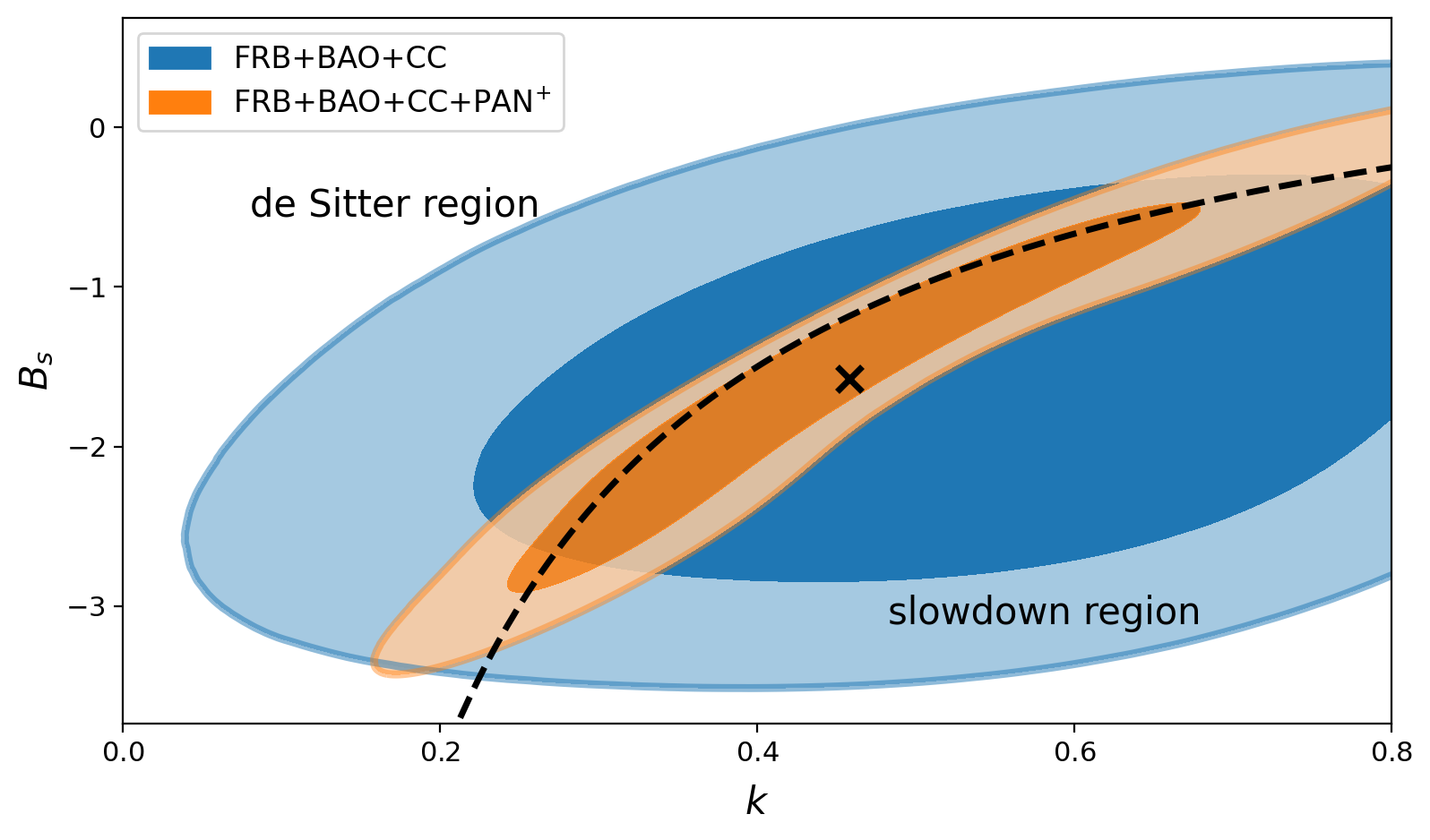} 
\qquad 
\includegraphics[width=0.455\linewidth]{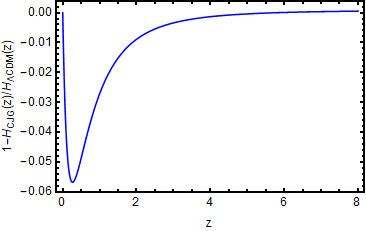}
\caption{Left: $B_s - k$ plane for the joint analysis, alongside  the curve $B_S=-\frac{k'}{k}$ (black dashed curve), separating regions $\cal{N}_3$ and $\cal{N}_4$. Right: plot of the relative deviation between the Hubble function of CJG and that of the $\Lambda$CDM model, $1-H_{CJG}(z)/H_{\Lambda \text{CDM}}$, using the best-fit parameters from (b) FRB + BAO + CC + $\rm PAN^{+}$.}
\label{Ez}
\end{figure*}


A complementary analysis is provided  by the statefinder parameters  $r$ and $s$, which account for the deviations of the CJG model from $\Lambda$CDM. In Fig. \ref{weffq} (left), the trajectory traced by the CJG  in the $s-r$
plane is shown, obtained by parametrically plotting  Eqs. (\ref{rz}) and (\ref{sz}) with varying $z$. If the role of DE is played by a cosmological constant, the value of $r$ remains constant at $r=1$ throughout the matter-dominated epoch and into the future (for $z \lesssim 10^4$).  In the $s-r$ plane, the fixed point $\{s,r\}= \{0,1\}$ corresponds to the $\Lambda$ CDM case. For the CJG case, at past times ($z>0$), the pair $(s,r)$  is $\sim (0,1)$, indistinguishable from $\Lambda$CDM. However,
 currently, the parameters are $s\simeq 8$ and $r\simeq -6$, indicating a deviation of the CJG trajectory from  $\Lambda$CDM. In Fig. \ref{weffq} (right) the effective EoS parameter  $w_{eff}$ (\ref{weff}) is displayed. The solid-black and solid-blue lines correspond to the CJG and $\Lambda$CDM models, respectively. The effective EoS parameter  $w_{eff}$  provides a summary of the combined effects of various components in the universe. As observed, $w_{eff}$ changes during the matter-dominated and dark-energy-dominated epochs, reflecting the dominant component's EoS parameter at that time. For CJG and $\Lambda$CDM, the effective EoS parameters are practically indistinguishable in the past. However, around the present epoch, in CJG case, $w_{eff}$ starts to increase and at some point in the future  it becomes positive, indicating cosmic deceleration.  This corroborates the late-time behavior of CJG initially observed from the deceleration parameter 
(Fig. \ref{weffq}).
 
\begin{figure*} [h!]
\includegraphics[width=0.48\linewidth]{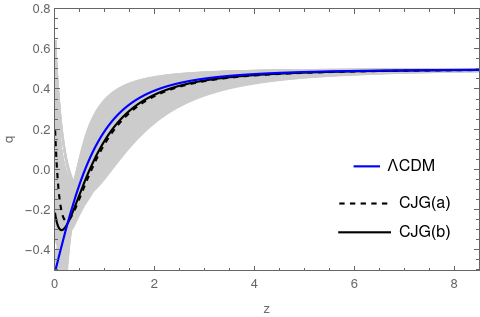} \qquad 
\includegraphics[width=0.48\linewidth]{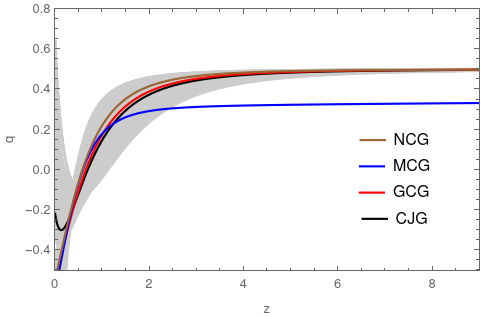} \caption{Deceleration parameter as a function of redshift for the best-fit values found in the joint analysis (b) FRBs + BAO + CC + PAN$^+$ (continuous black curve) alongside its 1$\sigma$ region, together with the best-fit for (a) FRBs + BAO + CC (dashed black curve) and} $\Lambda$CDM case (left) and other Chaplygin-type cases (right).
\label{figure4}
\end{figure*}


\begin{figure*}
\includegraphics[width=0.48\linewidth]{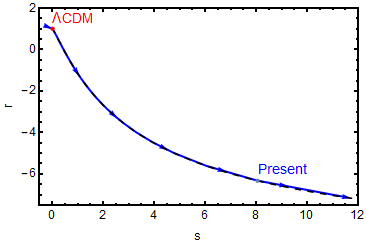}
\includegraphics[width=0.5\linewidth]{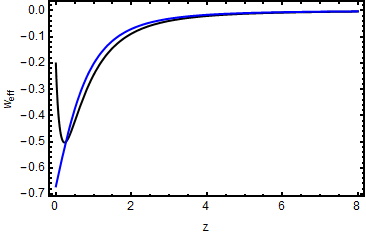}
\caption{Plots of the trajectory on the $s-r$ plane for the statefinder analysis (left panel) and the evolution of the effective EoS parameter against $z$ for the CJG cosmological model. For both plots, we have used the best fit parameters found in the analysis (b) FRB + BAO + CC + $\rm PAN^{+}$.}
\label{weffq}
\end{figure*}


\section{Conclusions and final remarks}\label{conclu}

We have explored an  extension of the Chaplygin EoS, derived from a specific generalization of the Chaplygin inflationary model within the framework of the Hamilton-Jacobi formalism and elliptic functions \cite{Rengo1}. By imposing conditions to obtain physical solutions for the energy density, we reduced the parameter space to different sub-regions where the model is free of singularities within the interval $0 < a < 1$, and the density remains always positive. Interestingly, two of these sub-regions are compatible with the possibility of a new deceleration phase in the future, following the current acceleration.  Then, we used different data samples such as SNIa, CC, FRBs and BAO to investigate the viability of the CJG cosmological model including radiation and matter. Constraints on the free parameters we obtained using a  statistical MCMC analyses and  their error bars are presented in Table \ref{table1}. Best-fit values for joint analyses (a) and (b)}support: the cosmic slowing down phenomenon (see Fig.~{\ref{Fig3}}) with  a maximum of acceleration at $z\sim 0.02$, $\alpha$ values such that $\alpha+1>0$ and negative values for $B_s$, with  a $\chi^2_{\nu} \sim 1$.  The preference for negative values for the  $B_s$ parameter  implies a lack of direct correspondence with the GCG. Additionally, the best-fit values result in a cosmology almost indistinguishable from 
$\Lambda$CDM in the past, deviating from it only in recent epoch, and diverging from the eternal dS expansion.

Exploring CJG expands two families of models. Firstly, it increases the number of models within the Chaplygin type family. Unlike other models in this family, however, CJG  allows for a future acceleration-deceleration transition. Secondly, it enlarges the family of models that predict  future deceleration, such as interacting models, $w$ parametrizations, decaying dark matter, extensions of  quintessential cosmology, and others (see, for example, \cite{Shafieloo2009, Vargas2011, Magana2014, Escobal2023}, and references therein). Moreover, it has been observed that a potential future transition to a decelerating stage, thereby naturally ending the eternal accelerating regime, is significant from a physical perspective \cite{Escobal2023}. An eternal de Sitter phase disagrees with the requirements of the S-matrix describing particle interactions and can be problematic within the framework of string theory \cite{Fischler2001, Cline2001, Hellerman2001}. Our model comparison with $\Lambda$CDM yields $\Delta \chi^2_{\text{min}} \sim 6.4$ for case (a) and $\Delta \chi^2_{\text{min}} \sim 28.5$ for case (b), indicating strong and very strong evidence for $\Lambda$CDM, respectively. Regarding the information criteria, $\Delta$AIC $\sim 6.4$, and $\Delta$BIC $\sim 18.7$ for case (a). Following Jeffreys' scale \cite{Liddle2007}, this suggests strong evidence ($\Delta$AIC) and very strong evidence ($\Delta$BIC) for $\Lambda$CDM. For case (b), we find $\Delta$AIC $\sim 30.5$, and $\Delta$BIC $\sim 50.7$, providing very strong evidence against CJG. However, it is important to emphasize that these results are inherently model-dependent. A definitive understanding of the universe’s ultimate fate may require a model-independent, fully data-driven approach, opening opportunities for future research.



\section*{Acknowledgements}
J.A.S.F. thanks FAPES  for the fellowship and financial support (045/2024 - P: 2024-M88VK). Also, J.A.S.F. is grateful for the hospitality of the Instituto Argentino de Radioastronomia (IAR), where part of this work was carried out. W.S.H.R. thanks FAPES (PRONEM No 503/2020) for the financial support under which this work was carried out. N. V. acknowledges the financial support of Fondecyt Grant 1220065. 
J. R. V. is partially supported by Centro de F\'isica Te\'orica de Valpara\'iso (CeFiTeV). N. V. and J. R. V. thank the CEUNES/UFES for their warm hospitality while some of this work was carried out.

\bibliographystyle{ieeetr}
\bibliography{Biblio_v1}
\appendix
\section{Sub-region for cosmic slowing down}
In this appendix, we present a detailed computation of the condition for having cosmic slowing down of the current acceleration. First, we consider the EoS parameter for the CJG model as a function of redshift
 \begin{eqnarray}
\omega (z) = -1 + \frac{(1-B_s)k}{k(1-B_s)+(k'+B_s k)(1+z)^{-3(1+\alpha)}} + \frac{(1-B_s)k'}{(k'+B_s k)(1+z)^{-3(1+\alpha)}-(1-B_s)k'} \,.
\end{eqnarray}
Next, we use the change of variable $y=\rho^{1+\alpha}$ in Eq.~(\ref{omega}). The EoS parameter and its derivative as functions of the new variable are
\begin{eqnarray}
w = -\frac{B_s k}{y} - 2k'+ \frac{k'}{B_s}y \,, \qquad \frac{d w}{dy} = \frac{B_s k}{y^2} +\frac{k'}{B_s} \,.
\end{eqnarray}
Considering the $y$ variable, the condition for cosmic deceleration (\ref{condition}) can be written as 
\begin{eqnarray}
\frac{d w}{dz} = \frac{d w}{dy}  \frac{d y}{dz}  <0 \,. 
\end{eqnarray}
For $B_s <0$, $\frac{d w}{dy}$  is always negative, so $\frac{d y}{dz}$ must be always positive. In other words,
\begin{eqnarray} \label{cond}
\frac{d y}{dz} =3(1+\alpha)B_s (1-B_s)\frac{(k'+B_s\,k)(1+z)^{-1-3(1+\alpha)}}{[(k'+B_s\,k)(1+z)^{-3(1+\alpha)}-(1-B_s)k']^2} >0  
\end{eqnarray}
Then, this last condition reduces to
\begin{eqnarray}
(1+\alpha)(k'+B_s\,k) <0 \,. 
\end{eqnarray}
In the case $\alpha +1 \ge 0$, this is valid for $B_s < -\frac{k'}{k}$. 

For  $B_s > 0$, $\frac{d w}{dy}$  is always positive, and then  condition (\ref{cond}) reduces to
\begin{eqnarray}
(1+\alpha)(1-B_s) <0  
\end{eqnarray}
$ \alpha +1 <0$, this condition is  valid when $B_s>1$. Therefore , the CJG  presents cosmic slowdown in ${\cal{N}}_1$ and ${\cal{N}}_4$.

\end{document}